\documentclass[fleqn,twoside]{article}
\usepackage{espcrc2}
\usepackage{graphicx}
\usepackage{epsfig}
\usepackage{latexsym}
\newcommand{\Dslash}{\slash\hspace{-2.5mm}D}

\title{
		{\vspace{-1.2em} \parbox{\hsize}{\hbox to \hsize
		{\hss  \normalsize TRINLAT-03/05}}} \\
	Heavy hadrons on an anisotropic lattice}

\author{Justin Foley\address[TCD]{TrinLat Collaboration \\ 
        School of Mathematics, Trinity College, Dublin 2, Ireland},
	Alan \'O Cais\addressmark[TCD], Mike Peardon\addressmark[TCD] 
	and Sin\'ead M. Ryan\addressmark[TCD]
	\thanks{talk presented by Sin\'ead Ryan}
	}

\begin{document}

\begin{abstract}
Results from simulations of quarkonia and heavy-light mesons on an 
anisotropic lattice are presented. The improved quark action and 
action-parameter tuning used in this study are discussed.
\end{abstract}

\maketitle

\section{DESIGNING A 3+1 ACTION}
A fermion action, which is useful for simulations at extreme anisotropies and 
has discretisation errors of ${\cal O}(\alpha_sa,a_s^3,a_t^2)$, is constructed. 
The details of this action are in Refs~\cite{ref:FOPR-in_prep,ref:MJP-cairns}.
Our starting point is the usual isotropic rotation, used in the Sheikholeslami-Wohlert (SW) and 
D234 actions
\begin{eqnarray}
  \psi       &=& [1-\frac{ra}{4}(\Dslash -m)]\psi^\prime ,\nonumber\\
  \bar{\psi} &=& \bar{\psi^\prime}[1-\frac{ra}{4}(\Dslash -m)] .
  \label{eqn:iso_rotation}
\end{eqnarray}
For a fixed Wilson parameter, $r$ (usually equal to one) doublers may reappear as 
$a_t$ is made small, which 
hinders simulations with $a_t\ll a_s$. The solution proposed here is to treat the temporal and 
spatial directions differently. The temporal rotations are retained since such
one-hop improvement terms ensure the transfer matrix is positive definite and no 
``ghost'' states can propagate. Thus Eqs.~\ref{eqn:iso_rotation} are modified to read
\begin{eqnarray}
  \psi       &=& [1-\frac{ra_t}{4}(\gamma_0D_0 -m)]\psi^\prime , \nonumber\\
  \bar{\psi} &=& \bar{\psi^\prime}[1-\frac{ra_t}{4}(\gamma_0D_0 -m)] 
  \label{eqn:aniso_rotation}
\end{eqnarray}
and once the covariant temporal derivative is discretised the temporal doublers are 
removed. 
The spatial doublers remain and are removed through the addition of an 
(irrelevant) higher-dimensional operator, a $D^4_i$ term, to the Dirac operator. The resulting
improved action can then be written
\begin{eqnarray}
S^\prime = \bar{\psi^\prime}M_r\psi^\prime -\frac{ra_t}{2}\bar{\psi^\prime}
           \left( D_t^2 - \frac{g}{2}\epsilon_iE_i\right)\psi^\prime  \nonumber \\
            + sa_s^3\bar{\psi^\prime}\sum_iD^4_i\psi^\prime ,
	\label{eqn:continuum-action}
\end{eqnarray}
where $M_r=\mu_r\gamma_iD_i + \gamma_0D_t + \mu_rm$ and $\mu_r=(1+\frac{1}{2}ra_tm)$. 
The Wilson-like parameter, $s$ is chosen such that the doublers receive 
a sufficiently large mass. A lattice discretisation scheme is now quite simple. 
An improved discretisation is only needed for the 
$\gamma_iD_i$ term in Eq.~\ref{eqn:continuum-action} since the 
simplest discretisation would lead to ${\cal O}(a_s^2)$ errors. Including the 
gauge fields and the mean-link improvement coefficients, $u_s$ and $u_t$ the 
lattice fermion matrix reads
\begin{eqnarray}
\hspace{-5ex}&&\hspace{-3ex}M_L\psi_x=\hspace{-.5ex}\frac{1}{a_t}\left\{\left(\mu_rma_t+\frac{18s}{\xi}+r+\frac{ra_t^2g}{4}\epsilon_iE_i\right)
		\psi_x \right. \nonumber \\
	\hspace{-2ex}&&\hspace{-4ex}\left.-\frac{1}{2u_t}
			\left[ (r-\gamma_0)U_t(x)\psi_{x+\hat{t}} +
			(r+\gamma_0)U_t^\dagger(x-\hat{t})\psi_{x-\hat{t}}\right] 
			\right. \nonumber \\
	\hspace{-2ex}&&\hspace{-2ex} \left. -\frac{1}{\xi}\sum_i\left[ 
			\frac{1}{u_s}(4s-\frac{2}{3}\mu_r\gamma_i)U_i(x)\psi_{x+\hat{\imath}} 
			\right.\right. \nonumber \\
	\hspace{-2ex}&&\hspace{-2ex} \left. \left.+\frac{1}{u_s}(4s+\frac{2}{3}\mu_r\gamma_i)
			U_i^\dagger(x-\hat{\imath})\psi_{x-\hat{\imath}}  \right.\right.\nonumber \\
	\hspace{-2ex}&&\hspace{-2ex}\left.\left.-\frac{1}{u_s^2}(s-\frac{1}{12}\mu_r\gamma_i)
			U_i(x)U_i(x+\hat{\imath})\psi_{x+2\hat{\imath}}  \right.\right. \\ 
	\hspace{-2ex}&&\hspace{-2ex} \left.\left.-\frac{1}{u_s^2}(s+\frac{1}{12}\mu_r\gamma_i)
			U_i^\dagger (x-\hat{\imath})U_i^\dagger (x-2\hat{\imath})
			\psi_{x-2\hat{\imath}}\right] \right\}. \nonumber
\end{eqnarray}
Choosing $r=1$ and $s=1/8$ yields the sD34 action proposed in 
Ref~\cite{ref:Hashimoto-Okamoto}. The 
authors investigated the radiative corrections to this action and at 
one-loop in perturbation 
theory find no contribution from ${\cal O}(\alpha_sa_sm_q)$ terms, which would spoil simulations 
with heavy quarks. This is a key advantage of this action over anisotropic SW- or D234-type 
actions where such terms may appear, 
depending on the choice of the parameter, $r$~\cite{ref:ask_etal}.
\section{RESULTS}
\begin{figure}[ht]
	\centering
	\epsfxsize=7.5cm\epsfbox{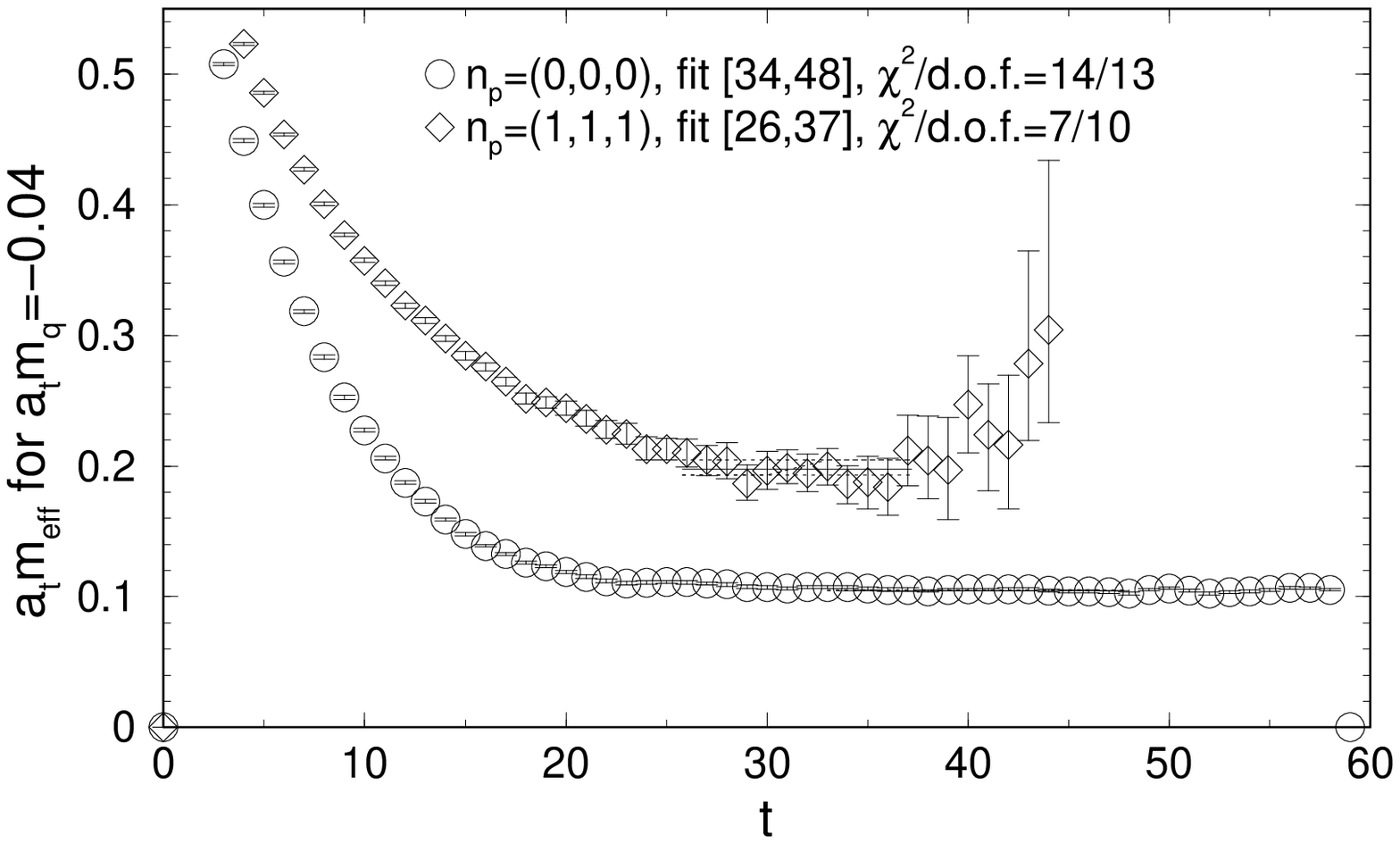}\\
	\epsfxsize=7.5cm\epsfbox{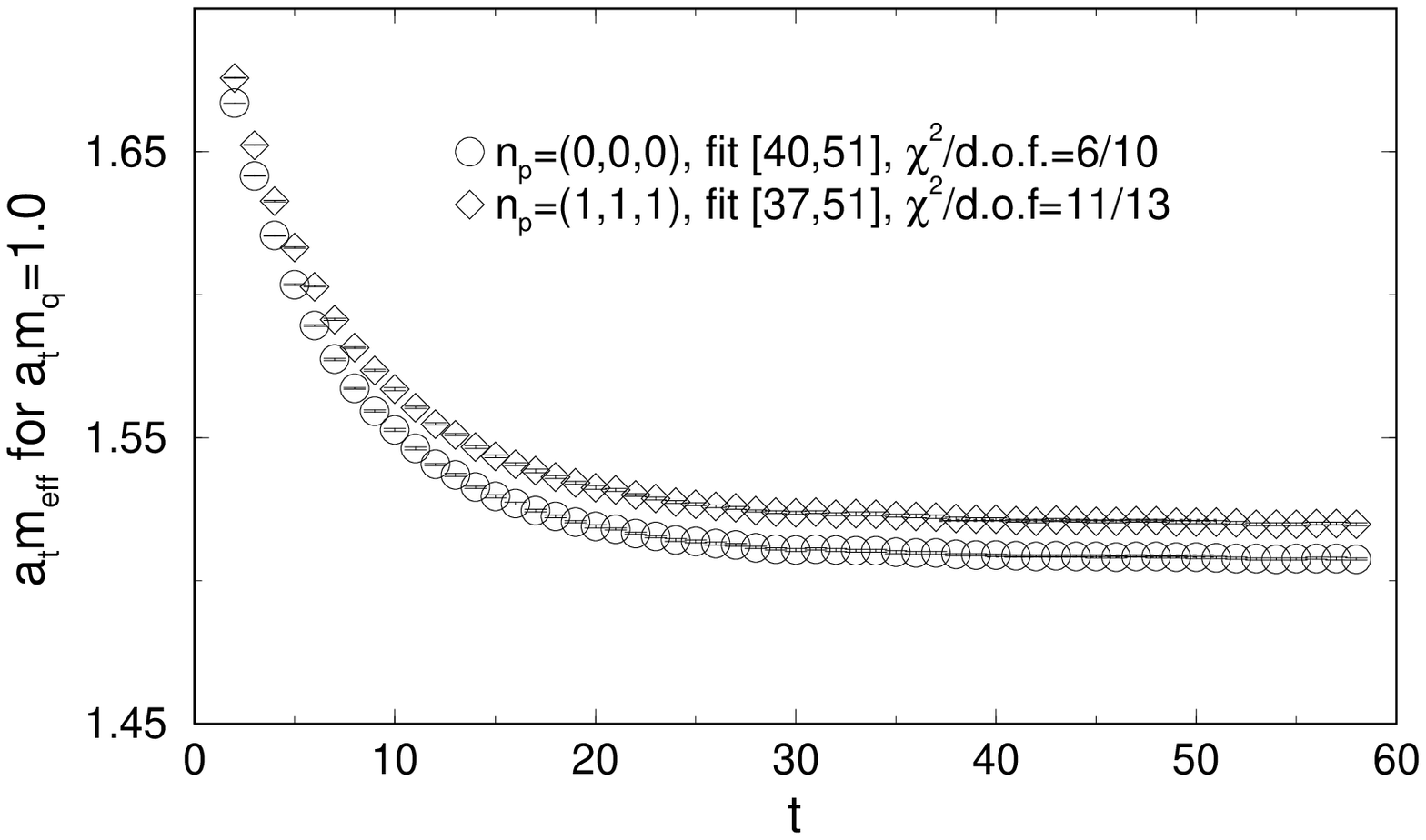}
	\vspace{-1cm}
	\caption[fig:PSm0.04_1.0]{Effective mass plots $a_tm_q=-0.04,1.0$ respectively. For $a_tm_q=-0.04, 
	n_p=(1,1,1)$ the signal becomes very noisy at $t\geq 50$.}
	\vspace{-4ex}
	\label{fig:PSm0.04_1.0}
\end{figure}

Our quenched simulations use an improved gluon action designed for precision 
glueball simulations on anisotropic lattices~\cite{ref:MorningstarPeardon}. 
We omit the temporal rotations in this exploratory 
study, leaving a leading discretisation error of ${\cal O}(a_t)$. Further details of 
the simulation and parameter values are in Table~\ref{tab:latt-details}. 
\vspace{-3ex}
\begin{table}[ht]
\begin{center}
\begin{tabular}{c|c}
\hline
\# gauge configs.  & 100 \\
Volume             & $10^3\times 120$ \\
$a_s$              & 0.21fm \\
$a_s/r_0$	   & 0.432 \\
$\xi=a_s/a_t$      & 6 $^\dagger$\\
\hline
\end{tabular}
\caption{Lattice details. $^\dagger$The gauge anisotropy was tuned nonperturbatively, from the static quark 
         potential.}
	\label{tab:latt-details}
\end{center}
\end{table}
\vspace{-6ex}
A range of masses is investigated from $a_tm_q=-0.04$, 
\footnote{This corresponds to a positive quark mass since Wilson-type actions have 
an additive mass renormalisation.} 
close to the strange quark mass, to $a_tm_q=1.5$. 
Data were accummulated at spatial momenta $(0,0,0),(1,0,0),(1,1,0)$ and $(1,1,1)$ in 
units of $2\pi /a_sL$, averaging over equivalent momenta. Degenerate and nondegenerate combinations 
of quark propagators are considered. The nondegenerate combination is made with the lightest quark, 
and each of the heavier quarks. These mesons are denoted heavy-light while the 
degenerate combination is heavy-heavy. 
Effective masses for both pseudoscalar (PS) and vector (V) particles were determined from 
single cosh fits. For all masses and momenta good fits to long plateux are 
observed. Fits to the lightest and heaviest PS masses are shown in 
Fig.~\ref{fig:PSm0.04_1.0}. 
\subsection{Mass dependence of $\xi$}
Quantum fluctuations renormalise the anisotropy so the input parameter, 
$\xi$ in the action is not the same as the anisotropy determined from a physical 
observable in the simulation. The parameter in the action must be ``tuned'' such that 
the measured anisotropy takes its required value.
The aim of this study was to investigate the mass dependence of the renormalised 
anisotropy, examining the extent to which the tuning must be repeated as $a_tm_q$ 
is varied. Both the precision of the determination 
and its deviation from the required anisotropy are of interest. 
We also examine the difference between the anisotropy determined from PS and V mesons. 

\begin{figure}[ht]
 	\epsfxsize=3cm\epsfbox{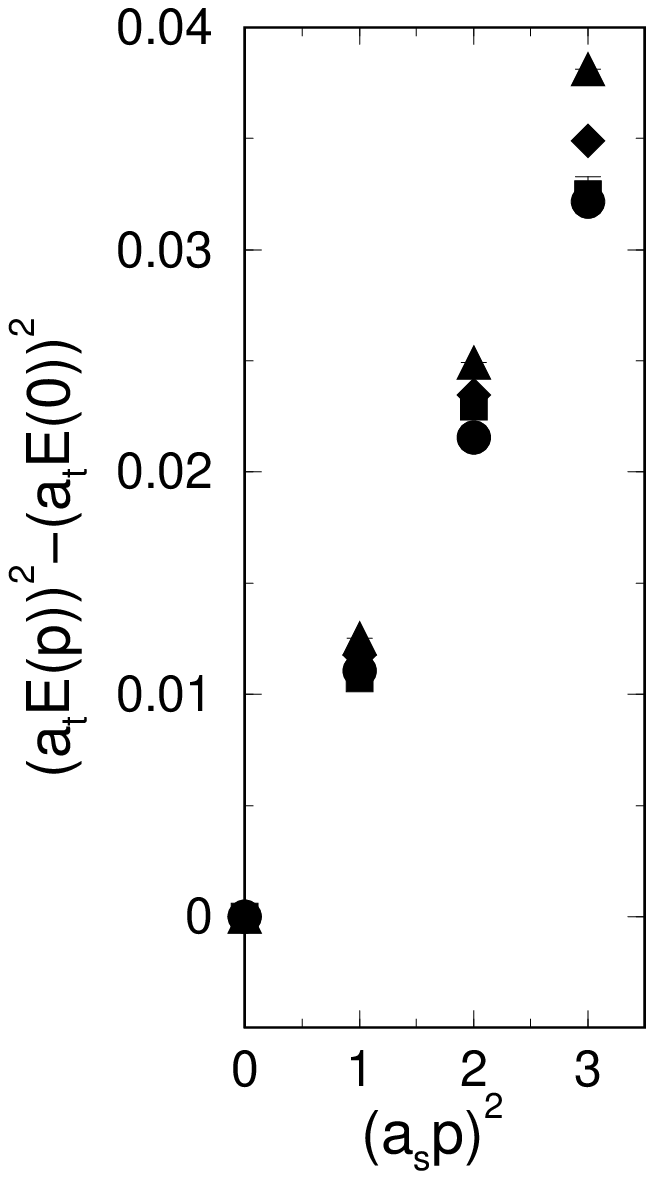}\vspace{-.5cm}
	\epsfxsize=3cm\epsfbox{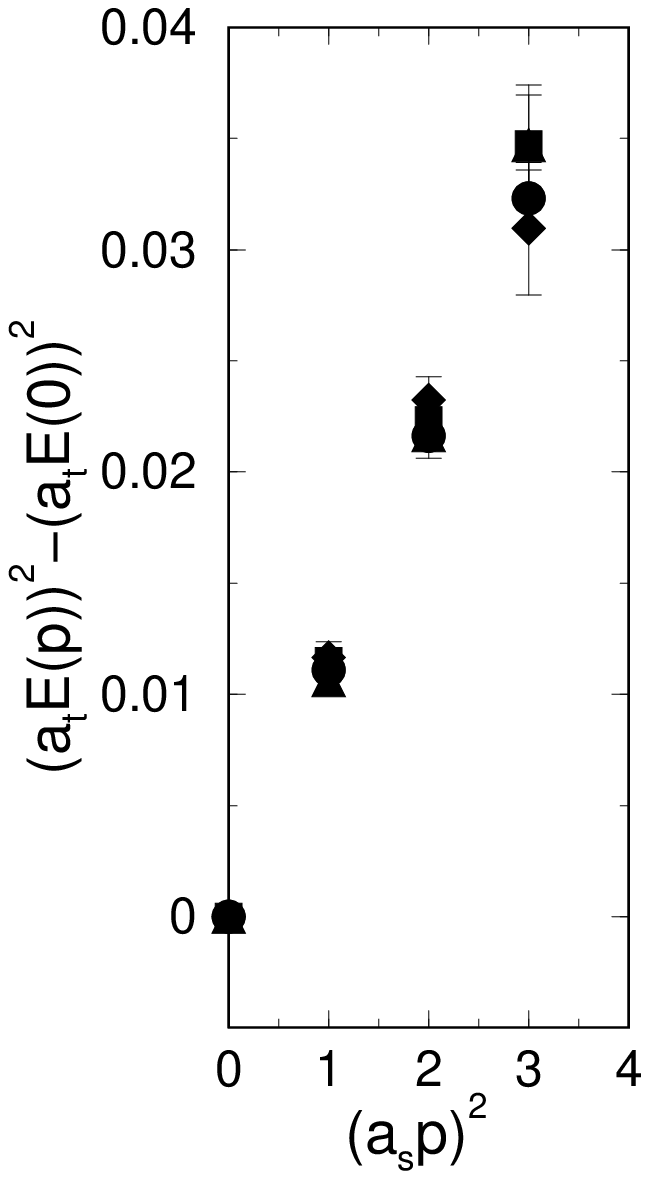}\vspace{-.5cm}
	\caption[fig:boosted-aniso]{Dispersion relations for heavy-heavy (left) 
				    and heavy-light (right)
	                            PS mesons. The symbols $\circ ,\Box ,\Diamond ,\triangle$ are 
				    $a_tm_q=0.1,0.2,0.5,1.0$ respectively.}
	\label{fig:disp-reln}\vspace{-5ex}
\end{figure}

\begin{figure}[ht]
	\centering
	\epsfxsize=7.5cm\epsfbox{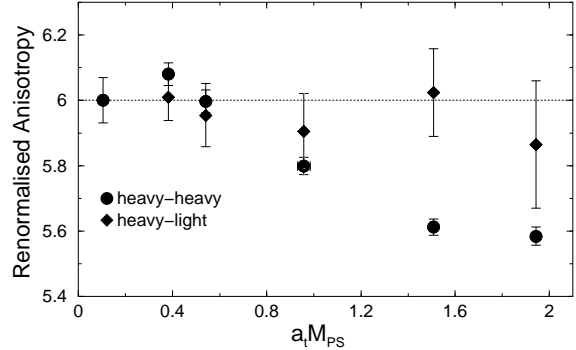}\vspace{-.5cm}
	\caption[fig:massdep-PSaniso]{The mass dependence of the anisotropy for PS mesons as a 
                                      function of the heavy-heavy meson mass in lattice units.}
	\label{fig:massdep-PSaniso}
\end{figure}

$\xi$ was tuned at the lightest mass by measuring the slope of the energy-momentum dispersion 
relation. A value of six from the dispersion relation requires 
$\xi=6.17$ in the action. The anisotropy was then computed, for a range of masses from their 
dispersion relations. Fig.~\ref{fig:disp-reln} is a representative 
sample of the dispersion relations for PS mesons. For clarity, $E^2(p)-E^2(0)$ is plotted 
as a function of $(a_sp)^2$. 
In general, very good linear (relativistic) behaviour is observed. 
The mass dependence of the measured anisotropy is shown in Fig.~\ref{fig:massdep-PSaniso} for 
PS mesons. Similar mass-dependence, although with larger statistical errors, is seen 
for the vectors. 
The plots show that up to $a_tm_q\sim 0.3$ the heavy-heavy and heavy-light determinations are in agreement. 
In addition the values are consistent, within errors, with the target anisotropy and are determined to 
$\approx 3\%$ or better accuracy. The charm quark mass on this lattice is close to $a_tm_q=0.2$, indicating 
that charm physics is feasible computationally and requires little parameter tuning. 

\begin{figure}[ht]
	\centering
 	\epsfxsize=2.8cm\epsfbox{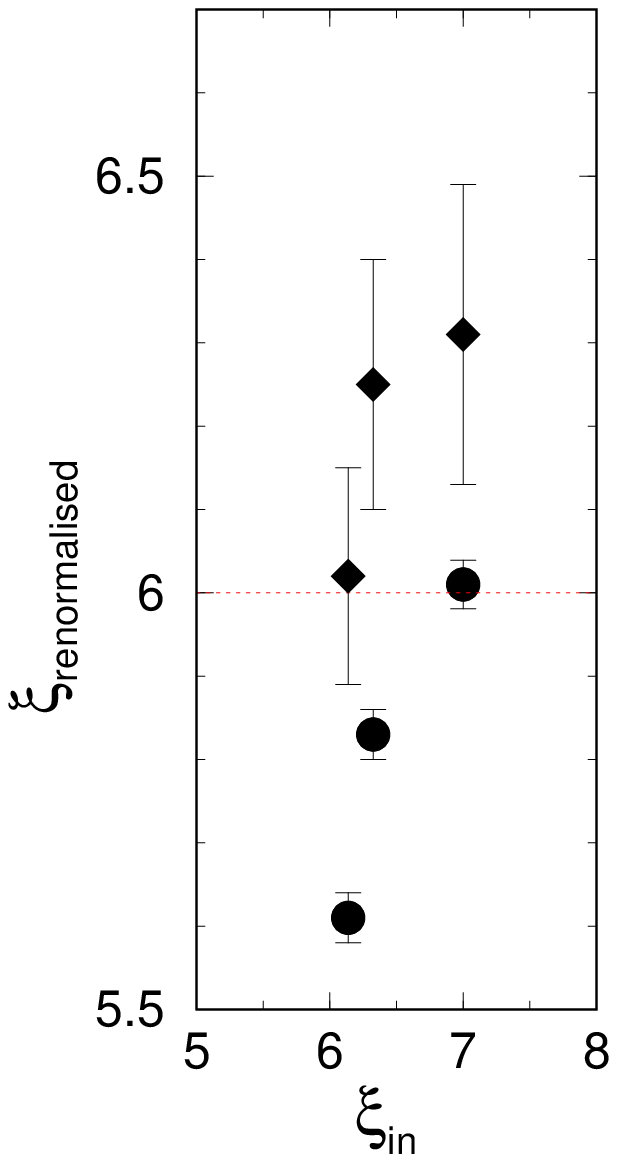}\vspace{-.5cm}
	\epsfxsize=2.8cm\epsfbox{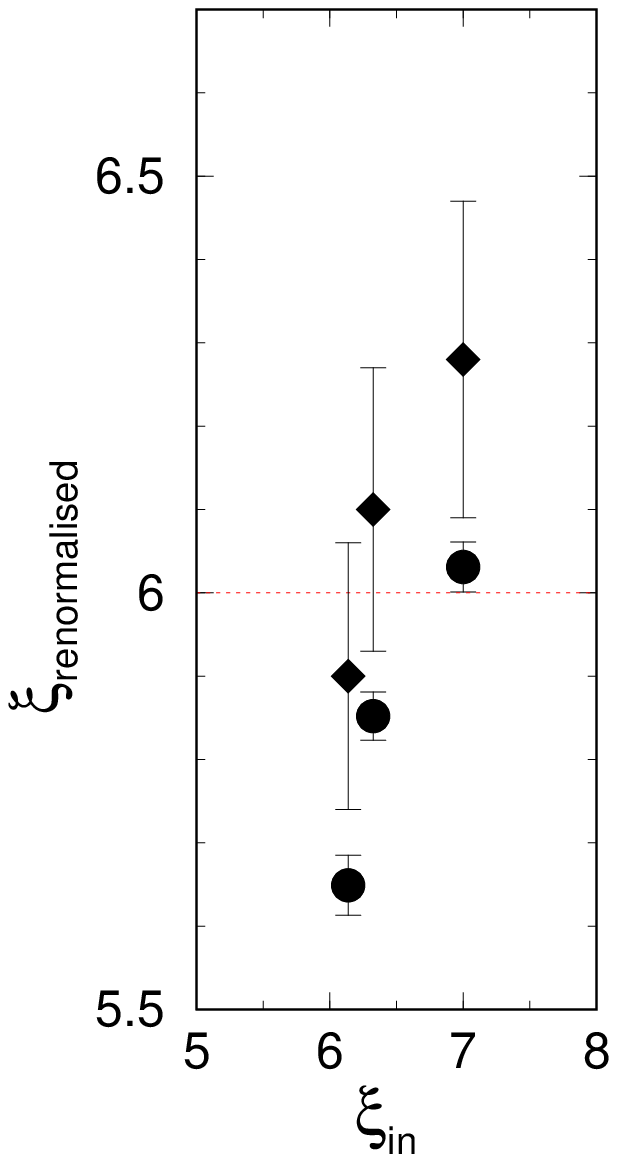}\vspace{-.5cm}
	\caption[fig:boosted-aniso]{The dependence of the renormalised anisotropy on the input parameter 
	                            $\xi$, for PS (left) and V mesons (right). $\circ$ and $\Diamond$ are 
				    heavy-heavy and heavy-light respectively.}
	\label{fig:boosted-aniso}\vspace{-4ex}
\end{figure}

For $a_tm_q>0.5$, the heavy-heavy and heavy-light dispersion relations do not give a consistent value 
of the anisotropy. This dependence of the measured anisotropy on the input parameter $\xi$ in the 
action was studied at fixed quark mass. The results, for both PS and V particles are shown 
in Fig~\ref{fig:boosted-aniso}. 
The value of $\xi$, determined from the heavy-heavy dispersion relation, moves closer to its 
target value of six while $\xi$ determined from the heavy-light physics moves away from this value. 
It is also interesting to note the agreement between determinations of $\xi$ from pseudoscalar 
and vector particles. The tuning, described above was carried out for pions and it 
is reassuring that although the vector particles have larger statistical errors they nevertheless yield 
a consistent pattern for the mass dependence of $\xi$.
\section{FUTURE WORK}
A paper with more details and additional quark masses is in preparation~\cite{ref:FOPR-in_prep}. We plan 
to investigate the feasibilty of precision $b$-physics using the anisotropic lattice described here. 

\end{document}